\documentclass{aa}
\usepackage{amssymb}
\usepackage{graphicx}
\usepackage{txfonts}
\usepackage{amsmath}	
\usepackage{multicol}
\usepackage{multirow}        
\usepackage{bm}		
\usepackage{pdflscape}	
\usepackage{color}
\usepackage{natbib}
\usepackage{chngpage}
\usepackage{lipsum}
\usepackage{makecell}
\usepackage[figuresright]{rotating}
\usepackage{subfigure}
\usepackage[switch]{lineno}
\usepackage{subfigure} 
\usepackage{float}
\usepackage{chngpage}
\usepackage{booktabs}
\usepackage{array}
\usepackage{lscape}
\usepackage{rotating}

\begin{document}
\linenumbers
\title{Estimating redshift distributions of Fermi blazars with multiple machine-learning models}

\author{Kerui. Zhu\inst{1}
          \and
            Jiaming. Chen\inst{1}
          \and
            Yonggang. Zheng\inst{3}
          \and
            Ruixin. Zhou\inst{4}
          \and
            Shan. Chang\inst{1}
           \and
            Li. Zhang\inst{1}
           \and
            Niansheng. Tang\inst{2}
          }

\offprints{Li. Zhang}

\institute{Department of Astronomy, Key Laboratory of Astroparticle Physics of Yunnan Province, Yunnan University, Kunming, 650091, People's Republic of China\\
              \email{lizhang@ynu.edu.cn}
         \and
               Yunnan Key Laboratory of Statistical Modeling and Data Analysis, Yunnan University, Kunming, 650091, People’s Republic of China\\
               \email{nstang@ynu.edu.cn}
         \and
               Department of Physics, Yunnan Normal University, Kunming, 650092, People's Republic of China
         \and
               Institute of Information, Yunnan University of Chinese Medicine, Kunming 650500,  People's Republic of China
              }

\date{Submitted; Received ; accepted }

\abstract
{Approximately half of the Fermi-LAT detected blazars lack spectroscopic redshift measurements, which limits population studies and investigations of the cosmological evolution of the gamma-ray blazar population.}
{We aim to develop a multi-model ensemble framework to systematically evaluate the performance of machine-learning models for blazar redshift estimation and to provide predictive redshift probability distributions with quantified uncertainties for all redshift-unknown blazars.}
{We construct an ensemble of 35 machine-learning regression models in the $\mathrm{1/(1+z)}$ space. Predictions from individual models are combined to build non-parametric redshift probability density functions (PDFs) for each source, allowing an explicit characterization of predictive uncertainty. The source-level PDFs are subsequently stacked to derive population-level redshift distributions.}
{Across the diverse models within our framework, we find comparable predictive performance, indicating a feature-limited performance plateau. We compile a catalog of predictive redshift distributions for 1,590 blazars without measured redshifts. The stacked PDFs recover the characteristic double-peaked redshift structure of blazars without using source class labels. For blazar candidates of uncertain type, the predicted redshift distribution exhibits a bimodal structure, suggesting a slightly larger contribution from FSRQ-like sources.}
{Our multi-model statistical framework achieves competitive predictive performance while retaining faint sources. We provide predicted redshift distributions with quantified uncertainties for 1590 blazars without measured redshifts, offering improved support for population and evolutionary studies. More broadly, this approach highlights the practical value of uncertainty-aware multi-model regression ensembling in astronomical data analysis.}


\authorrunning{Zhu et al.}

\maketitle
\nolinenumbers

\section{Introduction}\label{sec:intro}
Redshift is an essential physical quantity that reflects the distance to astrophysical sources and plays a critical role in studies of the intrinsic radiation properties of gamma-ray sources, the extragalactic background light (EBL), and the intergalactic magnetic field (IGMF).  Gamma-ray sources detected by Fermi Large Area Telescope (Fermi-LAT) at high Galactic latitudes ($|b| > 10^\circ$) have been dominated by blazars (e.g., \citealt{2009ApJS..183...46A, 2022ApJS..260...53A}), with the total number exceeding 3,500 by 2022 \citep{2023arXiv230712546B}. Gamma-ray emission from blazars is generally dominated by non-thermal continuum radiation and lacks prominent spectral line features, making it impossible to measure redshifts directly from gamma-ray observations. At present, redshifts are primarily determined through the identification of emission or absorption lines observed at other wavelengths (e.g., optical or UV), followed by multi-wavelength statistical association with gamma-ray sources.

The multiwavelength association process involves source catalogs at different wavelengths (e.g., the Roma-BZCAT catalog; \citealt{2009A&A...495..691M} and the WISE blazar catalog; \citealt{2014ApJS..215...14D}) as well as a variety of statistical techniques, such as Bayesian and likelihood-ratio methods (\citealt{2020ApJS..247...33A}), resulting in considerable complexity and inevitably introducing association uncertainties \citep{2025ApJS..281...53G}. Moreover, with the extension of exposure time, improved understanding of gamma-ray diffuse emission models, and continuous optimization of analysis pipelines, newly detected blazars tend to be fainter and exhibit weaker variability, making the reliable identification of their optical counterparts and subsequent spectroscopic observations increasingly challenging.

Only about half of gamma-ray blazars have reported redshift measurements \citep{2022ApJS..263...24A}, significantly compromising the redshift completeness of existing samples. Owing to intrinsic differences in their physical and radiative properties, flat-spectrum radio quasars (FSRQs) with strong emission lines (equivalent width $> 5\,\text{\AA}$; \citealt{1991ApJ...374..431S}) generally allow reliable redshift determination, whereas sources lacking redshift measurements are predominantly BL Lac objects (BL Lacs), whose weak emission lines are often overwhelmed by strong non-thermal continuum emission, as well as blazar candidates of uncertain type (BCUs) with limited multi-wavelength information. Developing reliable methods to estimate redshifts for these sources remains an essential open problem in gamma-ray blazar studies.

Based on gamma-ray observational data reported by Fermi-LAT, machine-learning regression models are employed to estimate the redshifts of blazars \citep{2021ApJ...920..118D,2022ApJS..259...55N,2023MNRAS.521.4156C, 2024MNRAS.527.6198G,2024MNRAS.530.2282D}. \cite{2021ApJ...920..118D} employs a Super Learner ensemble that integrates several  machine-learning algorithms to estimate the redshifts of blazars in the Fourth Catalog of Active Galactic Nuclei (AGN) Detected by the Fermi-LAT (4LAC). Using the 4LAC-DR2 dataset, \cite{2022ApJS..259...55N} introduces SLOPE regularization to further improve the performance of \cite{2021ApJ...920..118D}, while also investigating bias-correction techniques for the predicted redshifts. \cite{2023MNRAS.521.4156C} adopts a CatBoost-based pipeline to estimate the redshifts of active galactic nuclei in 4LAC-DR3.  All three methods achieve comparable performance on the test dataset, yielding a root mean square error (RMSE) of approximately 0.45 and a Pearson correlation coefficient of about 0.7 between the predicted and true redshifts. Nevertheless, this level of accuracy is insufficient for reliable per-source redshift predictions.

Consequently, recent studies have transitioned from single-value redshift estimation toward predicting redshift distributions and quantitatively assessing the associated uncertainties. \cite{2024MNRAS.527.6198G} employs deep neural network to model the redshifts of active galactic nuclei in 4LAC–DR3 and incorporates variational inference to quantify predictive uncertainties at the model level, thereby deriving redshift probability distributions. \cite{2024MNRAS.530.2282D} adopts the Hierarchical Correlation Reconstruction approach to construct conditional probability distributions between multi-dimensional observational features and redshift, enabling the reconstruction of the redshift distribution for sources lacking redshift measurements in 4LAC–DR3. Despite these advances, several key limitations hinder the construction of a robust and complete redshift catalog for the entire blazar population.

First, many previous modeling approaches relied heavily on a subset of features with substantial missing values, which in turn necessitated aggressive data-cleaning procedures and resulted in the exclusion of a large number of faint sources. This compromises their applicability to the redshift-unknown sample, which is predominantly faint. Furthermore, another significant issue arises from the modeling  parameter space. Most existing methods operate directly in redshift space ($z$-space), where the highly skewed and non-uniform intrinsic distribution of blazars can bias regression models toward lower redshift predictions \citep{2023MNRAS.521.4156C,2024MNRAS.527.6198G}. Studies have shown that constructing regression models in the transformed space $1/(1+z)$ can effectively mitigate distributional skewness and improve predictive stability \citep{2021ApJ...920..118D}.
In addition, predictions based on a single machine-learning model are often sensitive to model assumptions and training fluctuations, raising concerns about robustness. Thus, how to effectively integrate multiple models while providing reliable estimates of predictive uncertainty remains an open question.

To address these issues, we propose a new framework for redshift distribution estimation based on an ensemble of machine-learning regression models. Our framework is designed for the full blazar sample and explicitly addresses distributional skewness by performing regression in the $1/(1+z)$ space. We employ multiple, diverse machine-learning models to generate independent redshift predictions for each source.

These predictions are treated as an ensemble of independent redshift realizations, from which we construct a non-parametric redshift probability density function (PDF) for each source and compile a comprehensive redshift distribution catalog for the entire sample.
The resulting source-level PDFs are subsequently aggregated to infer the redshift distributions of different blazar subclasses and to facilitate population-level studies.
By retaining low signal-to-noise sources and leveraging the ensemble of multiple machine learning models, we provides a robust and reliable reconstruction of redshift distributions without relying on aggressive sample cleaning.

The structure of this paper is organized as follows. Section \ref{sec:sample} describes the data sets, feature parameters, and preprocessing procedures used in this work. Section \ref{sec:model} provides a brief overview of the machine-learning methods adopted for redshift modeling. Section \ref{sec:comparison} presents a systematic comparison of the predictive performance of the different regression models. In Section \ref{sec:results}, we present the construction of redshift probability distributions for individual sources as well as for different blazar subclasses. Finally, Section \ref{sec:discussion} offers conclusion and discussion.

To ensure the reproducibility of our work, we fixed the random seed to ``123'' in the code involving random processes.

\section{Sample} \label{sec:sample}

\begin{figure*}
\centering
\begin{minipage}[t]{0.72\textwidth}
\vspace{0pt}
\includegraphics[width=\linewidth]{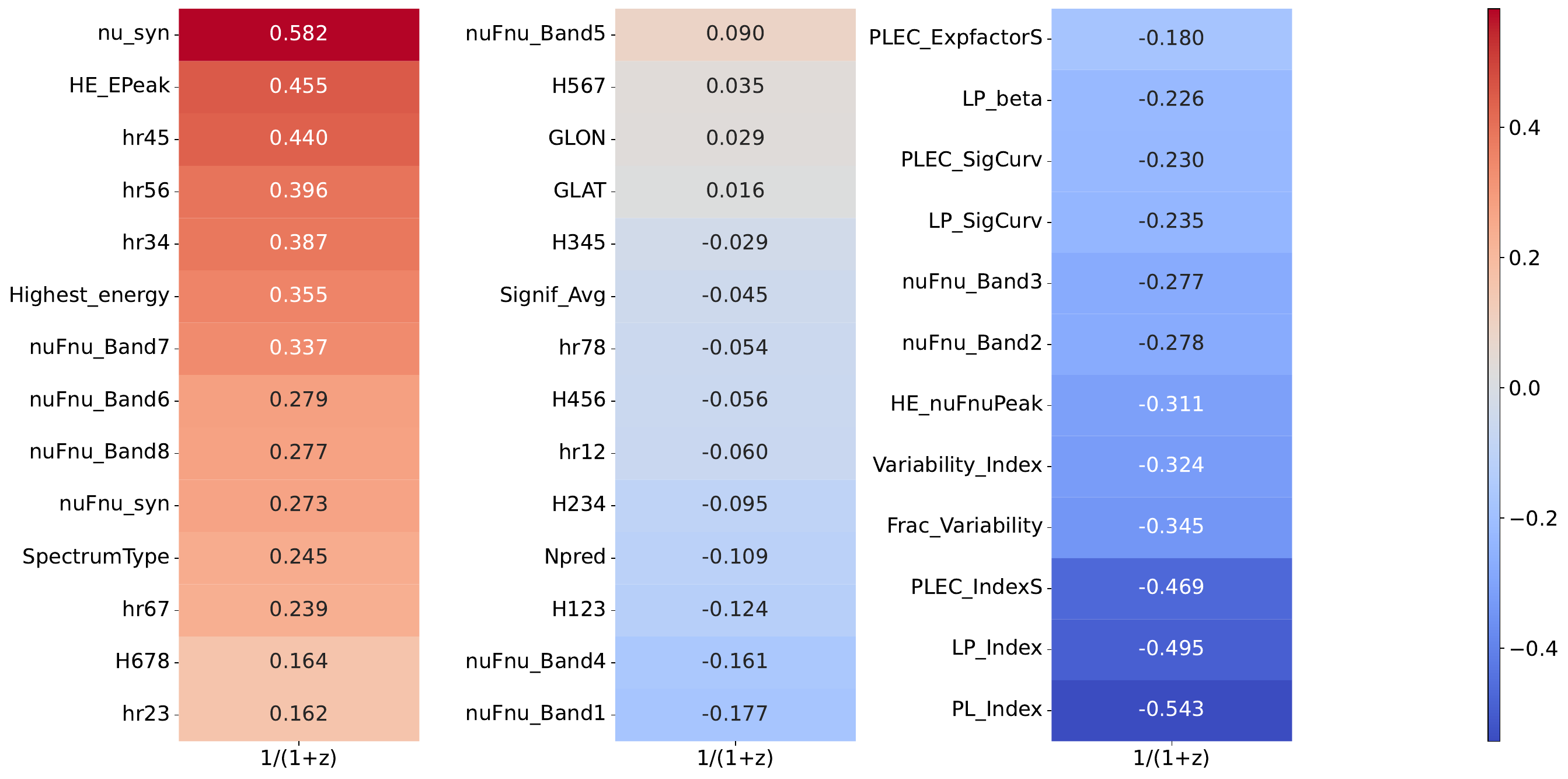}
\end{minipage}
\hfill
\begin{minipage}[t]{0.25\textwidth}
\vspace{0pt}
\caption{Correlation heat map of the 40 feature parameters with $1/(1+z)$. The map presents all feature parameters employed in this
study, and detailed parameter definitions can be found in \citep{2022ApJS..260...53A}. Numbers denote the Pearson correlation coefficients between each parameter and redshift, where red (blue) colors indicate positive (negative) correlations.}
\label{fig:1}
\end{minipage}
\end{figure*}

The latest Fermi-LAT AGN catalog, 4LAC-DR3, reports 3816 AGNs detected during observations conducted between 2008 and 2020 \citep{2022ApJS..263...24A}. We cross-matched the High-Galactic-Latitude sample of 4LAC-DR3\footnote{\url{https://fermi.gsfc.nasa.gov/ssc/data/access/lat/4LACDR3/table-4LAC-DR3-h.fits}} with the 4FGL-DR4 catalog\footnote{The 4FGL-DR4 catalog is described in \citealt{2023arXiv230712546B} and is publicly available at \url{https://fermi.gsfc.nasa.gov/ssc/data/access/lat/14yr_catalog/}.} and selected a final sample of 3333 blazars for this study. 
Among these sources, 1743 blazars have reported redshift measurements, including approximately 859 BL Lacs (hereafter BLLs, $\rm \bar{z}_{bll} =0.4240$), 753 FSRQs ($\rm \bar{z}_{fsrq} =1.1940$), and 131 BCUs ($\rm \bar{z}_{bcu} =0.7759$). The remaining 1590 sources lack redshifts and are predominantly composed of 523 BLLs and 1067 BCUs.

From the 4LAC-DR3 catalog, we extracted several physical parameters related to the broadband emission properties of the sources, including the synchrotron peak frequency, the synchrotron peak energy flux, the peak energy and corresponding energy flux of the high-energy component, as well as the energy of the highest-energy photon detected by Fermi-LAT. In addition, we obtained the detection significance, variability index, energy fluxes in individual energy bands, and spectral fitting parameters from the 4FGL-DR4 catalog. The complete list of adopted features is shown in Figure \ref{fig:1}, and detailed parameter definitions can be found in \cite{2022ApJS..260...53A}. 
Building on previous work, we incorporated two additional parameters: (1) hardness ratios ($\rm h_{ij}$) between energy bands as extended features \citep{2016ApJ...820....8S}, and (2) curvature coefficients ($\rm H_{ijk}$) among triplets of energy bands to quantify local spectral curvature \citep{2021ApJ...916...93Z}.
In total, 40 feature parameters were constructed and used in our analysis.

For feature parameters spanning more than two orders of magnitude, we first apply a logarithmic transformation to compress their dynamic range and mitigate the influence of outliers. All features are then normalized via Min-Max scaling to the $\rm [0,1]$ interval to suit neural-network-based models. Categorical parameters (e.g., SpectrumType) are encoded using a LabelEncoder implemented in \texttt{scikit-learn} \citep{Scikit-learn}.

Six of the features contain missing values: Frac\_Variability (missing rate 24.7$\%$), nu\_syn and nuFnu\_syn (22.7$\%$), HE\_nuFnuPeak and HE\_EPeak (15.2$\%$), and Highest\_energy (37.7$\%$). We find that missing values in these parameters are strongly associated with lower detection significance, indicating that the data are missing primarily for faint sources. To maximize the applicability of our models to the faint, redshift-unknown population, we retain all sources with missing values. We handle missing data differently depending on the algorithm's tolerance:

\begin{itemize}
\item[1.]
For models that are insensitive to missing values (e.g., tree-based models), we train and predict directly using the original, non-imputed feature set.

\item[2.]
For models that require complete data (e.g., neural networks), we first impute missing values using the K-nearest neighbors (KNN) method before model training and prediction.
\end{itemize}

The KNN-flag column in Table~\ref{tab:1} indicates whether a given model uses the KNN-imputed dataset.
To assess the robustness of the adopted missing-data treatment, we additionally performed a sensitivity analysis using Multiple Imputation by Chained Equations (MICE; e.g., \citealt{2021mlps.confE...1L,2022FrASS...936215G}). The comparison, based on several representative machine-learning models, indicates that the predictive performance is largely insensitive to the adopted imputation strategy.

Figure \ref{fig:1} presents the correlation heat map between the 40 feature parameters and $\rm 1/(1+z)$. We find that none of the features reaches the level of strong correlation, and only six parameters exhibit correlation coefficients greater than 0.4, corresponding to moderate correlations. Overall, the correlations between individual features and $\rm 1/(1+z)$ are generally weak, indicating that redshift information is limited when considering single parameters alone. This suggests that redshift dependence is primarily encoded through the joint and non-linear interactions among multiple features, which poses additional challenges for regression-based redshift estimation modeling.

\section{Machine learning model} \label{sec:model}

In this work, we construct and evaluate 35 machine learning regression models, which are grouped into five categories according to their underlying principles and model architectures.

\begin{table*}
\setlength{\abovecaptionskip}{0 cm}
\setlength{\belowcaptionskip}{0.1cm}
\begin{center}
\caption{Performance comparison of regression models for blazar redshift estimation}
\label{tab:1}
\centering
\small
\setlength{\LTleft}{0 cm} \setlength{\LTright}{0 cm} 
\begin{adjustwidth}{0cm}{0cm}
\resizebox{\textwidth}{!}{
\begin{tabular}{c c c c c c c}
\toprule
\textbf{No.} & \textbf{Method} & \textbf{KNN flag} & \textbf{Hyper-parameters}& \textbf{MAE} & \textbf{RMSE} & \textbf{Corr. Coef.} \\
\normalsize(1) & \normalsize(2) & \normalsize(3) &\normalsize(4) & \normalsize(5) &  \normalsize(6) &  \normalsize(7)      \\
\midrule
\multicolumn{7}{c}{\textbf{Linear methods}} \\
\midrule
1 & Linear Regression & Y&  & $0.1084\pm0.0037$ &$0.1376\pm 0.0045$&$0.6892\pm0.0207$ \\
2 & Ridge Regression & Y& $\alpha$: 1.0 & $0.1091\pm0.0037$ &$0.1376\pm 0.0044$&$0.6891\pm0.0212$ \\
3 & Lasso Regression & Y& $\alpha$: 0.0005 & $0.1108\pm0.0037$ &$0.1390\pm 0.0044$&$0.6816\pm0.0218$ \\
4 & Elastic-Net & Y& $\alpha$: 0.0005, L1/L2=0.5 & $0.1101\pm0.0037$ &$0.1383\pm 0.0044$&$0.6850\pm0.0215$ \\
5 & Bayesian Regression & Y& $\alpha_1$, $\alpha_2$, $\lambda_1$, $\lambda_2$: 0.0001 & $0.1089\pm0.0037$ &$0.1375\pm 0.0044$&$0.6896\pm0.0221$ \\
\midrule
\multicolumn{7}{c}{\textbf{Tree-based Methods}} \\
\midrule
6 & Decision Tree & Y& max\_depth: 5, min\_samples\_split: 2 & $0.1134\pm0.0037$ &$0.1461\pm 0.0045$&$0.6457\pm0.0233$ \\
7 & Random Forest & N & n\_estimators: 300, max\_depth: 10 & $0.1034\pm0.0033$ &$0.1326\pm 0.0042$&$0.7156\pm0.0221$ \\
8 & Extra Trees & Y & n\_estimators: 100, max\_depth: 10 & $0.1058\pm0.0035$ &$0.1344\pm 0.0044$&$0.7087\pm0.0215$ \\
9& Gradient Boosting & Y  & learning\_rate: 0.05, max\_depth: 2 & $0.1037\pm0.0036$ &$0.1331\pm 0.0045$&$0.7127\pm0.0213$ \\
10 & XGBoost & N & learning\_rate: 0.01, max\_depth: 5 & $0.1021\pm0.0034$ &$0.1310\pm 0.0043$&$0.7239\pm0.0204$ \\
11 & LightGBM & N & learning\_rate: 0.01, max\_depth: 5 & $0.1027\pm0.0033$ &$0.1318\pm 0.0043$&$0.7194\pm0.0200$ \\
12 & CatBoost & N & iterations: 500, max\_depth: 5 & $0.1058\pm0.0037$ &$0.1341\pm 0.0045$&$0.7104\pm0.0217$ \\
13 & AdaBoost & Y & n\_estimators: 500, max\_depth: 5 & $0.1044\pm0.0034$ &$0.1340\pm 0.0045$&$0.7085\pm0.0226$ \\
\midrule
\multicolumn{7}{c}{\textbf{Kernel methods}} \\
\midrule
14 & Kernel Ridge (RBF)& Y& $\alpha$: 0.001,$\gamma$: 0.005 & $0.1058\pm0.0038$ &$0.1349\pm 0.0044$&$0.7036\pm0.0205$ \\
15 & Kernel Ridge (Poly)& Y& degree: 5,$\alpha$:1 & $0.1087\pm0.0037$ &$0.1371\pm 0.0044$&$0.6923\pm0.0212$ \\
16 & SVR (RBF) &  Y & C: 1000, $\gamma$: 0.005 & $0.1050\pm0.0038$ &$0.1355\pm 0.0047$&$0.7059\pm0.0212$ \\
17 & SVR (Poly) & Y & C: 1, epsilon: 0.01, degree: 2 & $0.1048\pm0.0037$ &$0.1353\pm 0.0046$&$0.7065\pm0.0207$ \\
18 & Gaussian Process (Mat\'ern) & Y & $k(x, x') = (1.38)^2 \left(1 + \frac{\sqrt{3}|x-x'|}{14.4}\right) \exp\!\left(-\frac{\sqrt{3}|x-x'|}{14.4}\right)+ 0.0168 \, \delta_{x x'} $ & $0.1060\pm0.0037$ &$0.1350\pm 0.0043$&$0.7029\pm0.0205$ \\
19 & Gaussian Process (RQ) & Y & $k(x,x') =(0.649)^2\left(1 + \frac{|x-x'|^{2}}{2 \times 10^{5} \times 3.76^{2}}\right)^{-10^{5}}+ 0.0169\,\delta_{x x'} $ & $0.1056\pm0.0037$ &$0.1346\pm 0.0043$&$0.7051\pm0.0203$ \\
20 & Gaussian Process (RBF) & Y & $k(x,x') =(0.649)^2\exp\!\left(-\frac{|x-x'|^{2}}{2\times 3.76^{2}}\right)+ 0.0169\,\delta_{x x'} $ &  $0.1056\pm0.0037$ &$0.1346\pm 0.0043$&$0.7053\pm0.0202$ \\
\midrule
\multicolumn{7}{c}{\textbf{Neural Networks (4 Layers)}} \\
\midrule
21 & MLP-1 &  Y & layers: (128,64,32,16), activation: ReLU & $0.1013\pm0.0035$ &$0.1327\pm 0.0042$&$0.7181\pm0.0197$ \\
22 & MLP-2 &  Y & layers: (128,64,32,16), activation: Tanh & $0.1030\pm0.0037$ &$0.1352\pm 0.0050$&$0.7076\pm0.0240$ \\
23 & MLP-3 &  Y & layers: (128,64,32,16), activation: Sigmoid & $0.1016\pm0.0035$ &$0.1334\pm 0.0047$&$0.7163\pm0.0221$ \\
24 & MLP-4 &  Y & layers: (256,128,64,32), activation: ReLU & $0.1016\pm0.0038$ &$0.1335\pm 0.0048$&$0.7160\pm0.0221$ \\
\midrule
\multicolumn{7}{c}{\textbf{Neural Networks (5 Layers)}} \\
\midrule
25 & MLP-5 &  Y & layers: (256,128,64,32,16), activation: ReLU & $0.1011\pm0.0034$ &$0.1329\pm 0.0045$&$0.7180\pm0.0210$ \\
26 & MLP-6 &  Y & layers: (256,128,64,32,16), activation: Tanh & $0.1042\pm0.0035$ &$0.1366\pm 0.0047$&$0.7032\pm0.0224$ \\
27 & MLP-7 &  Y & layers: (256,128,64,32,16), activation: Sigmoid & $0.1022\pm0.0036$ &$0.1342\pm 0.0048$&$0.7132\pm0.0218$ \\
28 & MLP-8 &  Y & layers: (512,256,128,64,32), activation: ReLU & $0.1014\pm0.0036$ &$0.1333\pm 0.0047$&$0.7166\pm0.0218$ \\
\midrule
\multicolumn{7}{c}{\textbf{Neural Networks (7 Layers)}} \\
\midrule
29 & MLP-9 &  Y & layers: (1024,512,256,128,64,32,16), activation: ReLU &  $0.1012\pm0.0033$ &$0.1331\pm 0.0044$&$0.7171\pm0.0205$ \\
30 & MLP-10 &  Y & layers: (1024,512,256,128,64,32,16), activation: Tanh & $0.1065\pm0.0039$ &$0.1394\pm 0.0051$&$0.6915\pm0.0260$ \\
31 & MLP-11 &  Y & layers: (1024,512,256,128,64,32,16), activation: Sigmoid & $0.1036\pm0.0036$ &$0.1355\pm 0.0048$&$0.7068\pm0.0224$ \\
\midrule
\multicolumn{7}{c}{\textbf{Additional Methods}} \\
\midrule
32 & Tabpfn &  Y & version: 2.0 & $0.0991\pm0.0033$ &$0.1291\pm 0.0045$&$0.7332\pm0.0200$ \\
33 & FT-Transformer &  Y & num\_heads: 8, num\_attn\_blocks: 4 & $0.1166\pm0.0142$ &$0.1472\pm 0.0158$&$0.7026\pm0.0225$ \\
34 & TabNet &  Y & n\_d: 24, n\_a: 24, n\_steps: 3, gamma: 1.3, lambda\_sparse:0 & $0.1067\pm0.0038$ &$0.1355\pm 0.0047$&$0.7041\pm0.0214$ \\
35 & NODE &  Y & num\_trees: 512, depth: 6 & $0.1046\pm0.0038$ &$0.1349\pm 0.0048$&$0.7051\pm0.0224$ \\
\bottomrule
\end{tabular}}
\end{adjustwidth}
\end{center}
{\footnotesize{{Note. Column (3) indicates whether missing values are imputed using the KNN interpolation method. 
Column (4) lists the main Hyper-parameters adopted for each model. 
Columns (5)–(7) report the mean MAE, RMSE, and Pearson correlation coefficient, together with their corresponding standard deviations, obtained from 100 independent random train/test splits.
}}}
\end{table*}

\begin{figure*}
\centering
 \includegraphics[width=0.85\textwidth]{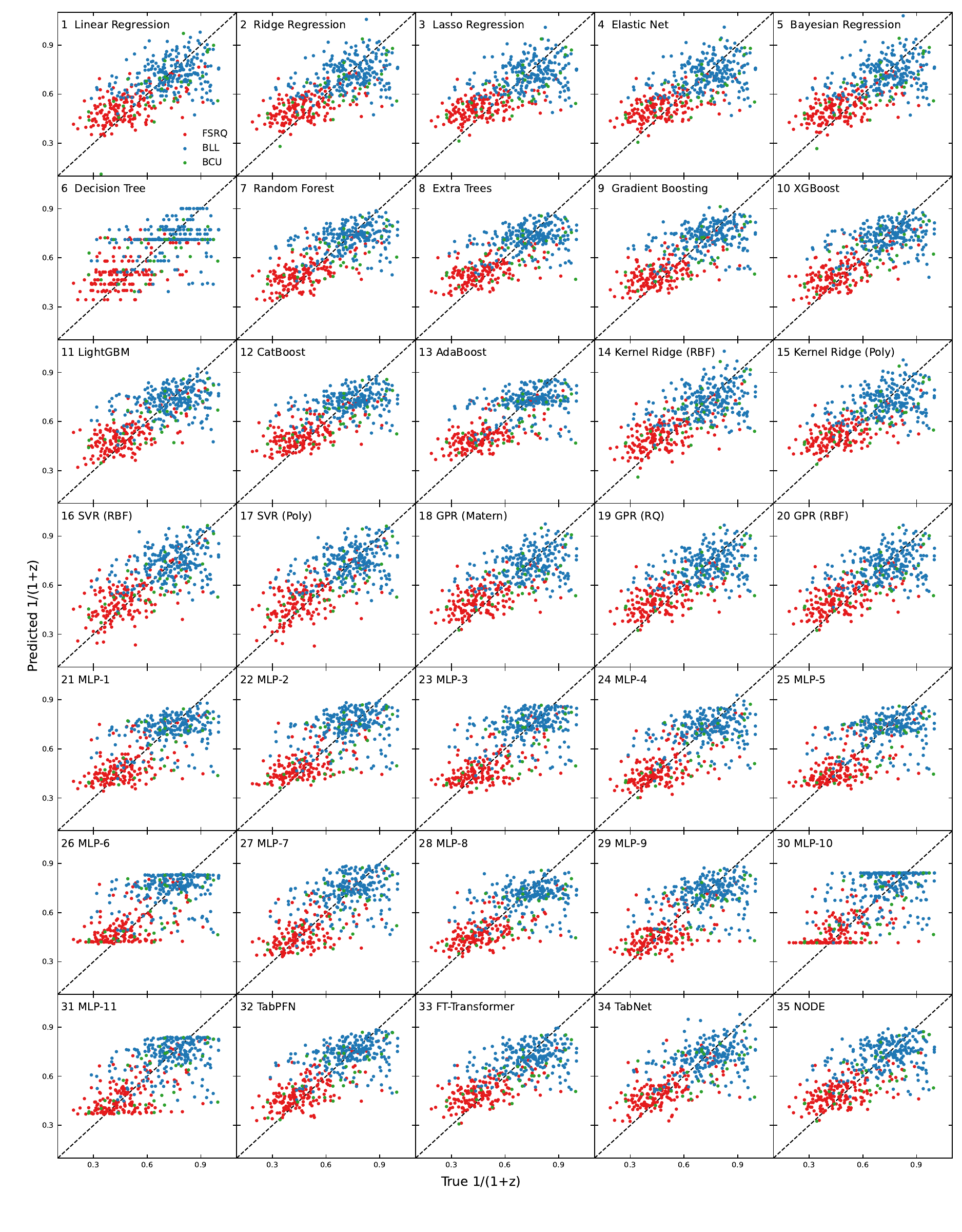}
\caption{Scatter relations between the predicted and true values of $\rm 1/(1+z)$ for the test set, shown for 35 different machine-learning regression models. Each panel corresponds to one regression model.
Red, blue, and green points denote FSRQs, BLLs, and BCUs, respectively. 
The black dashed line represents the ideal case in which the predicted values are equal to the true values.}
 \label{fig:2}
\end{figure*}

\subsection{Linear methods} 

We adopt linear regression models as baseline approaches to characterize linear relationships between the feature parameters and redshift. Five models are considered, including ordinary linear regression, Ridge regression \citep{Ridge}, Lasso regression \citep{LassoENet}, Elastic Net \citep{LassoENet}, and Bayesian regression \citep{Bayesian}. Ridge regression, Lasso regression, and Elastic Net incorporate different forms of regularization to mitigate multicollinearity and reduce over-fitting, while Bayesian regression introduces prior distributions on model parameters to account for uncertainty. Owing to their simplicity and interpretability, these models provide a useful benchmark for evaluating more complex methods.

\subsection{Tree-based methods} 

Tree-based models and ensemble learning methods are well suited for capturing non-linear relationships between feature parameters and redshift, while being relatively insensitive to feature scaling. In this work, we adopt a variety of tree–based regression models, including decision tree \citep{decisiontree}, Random Forest \citep{Randomforests}, Extremely Randomized Trees (Extra Trees; \citealt{ExtraTrees}), Gradient Boosting \citep{Gradient}, XGBoost \citep{XgBoost}, LightGBM \citep{LightGBM}, CatBoost \citep{CatBoost}, and AdaBoost \citep{AdaBoost}. Random Forest and Extra Trees reduce variance and improve generalization by aggregating multiple decision trees, whereas gradient boosting–based methods (Gradient Boosting, XGBoost, LightGBM, and CatBoost) employ additive modeling strategies to reduce bias and achieve strong performance on complex nonlinear problems. AdaBoost iteratively reweights training samples to emphasize difficult cases.

\subsection{Kernel methods} 

Kernel-based methods address nonlinear regression by implicitly mapping features into a high-dimensional space via kernel functions, where linear modeling becomes feasible. We employ three representative techniques: Kernel Ridge Regression (KRR; \citealt{KRR}), Support Vector Regression (SVR; \citealt{SVM, SVR}), and Gaussian Process Regression (GPR; \citealt{GPR}).

KRR extends ridge regression by incorporating the kernel trick with $\ell_2$ regularization, while SVR utilizes kernel functions under the maximum-margin principle for robust generalization. Both are implemented using the radial basis function (RBF) and polynomial kernels to capture different nonlinearities.

GPR adopts a Bayesian formulation by placing a Gaussian process prior over functions, $f(\mathbf{x}) \sim \mathcal{GP}\!\bigl(0,\, k(\mathbf{x}, \mathbf{x}')\bigr)$. We consider three commonly used covariance kernels, each of which includes an additive noise term $\sigma_n^2 \delta_{\mathbf{x}\mathbf{x}'}$ to account for noise:

\begin{itemize}

    \item Mat\'ern kernel:
    \begin{equation}\label{eq:1}
        k_{\mathrm{Mat\acute{e}rn}}(\mathbf{x},\mathbf{x}')
        = \sigma_f^2 
        \left(1 + \frac{\sqrt{3}\, r}{\ell}\right)
        \exp\!\left(-\frac{\sqrt{3}\, r}{\ell}\right)
        + \sigma_n^2 \delta_{\mathbf{x}\mathbf{x}'},
        \quad r = \lVert \mathbf{x}-\mathbf{x}' \rVert.
    \end{equation}

    \item Rational Quadratic (RQ) kernel:
    \begin{equation}\label{eq:2}
        k_{\mathrm{RQ}}(\mathbf{x},\mathbf{x}')
        = \sigma_f^2 
        \left(1 + \frac{\lVert \mathbf{x}-\mathbf{x}' \rVert^2}{2\alpha \ell^2}\right)^{-\alpha}
        + \sigma_n^2 \delta_{\mathbf{x}\mathbf{x}'}.
    \end{equation}
    
     \item RBF kernel:
    \begin{equation}\label{eq:3}
        k_{\mathrm{RBF}}(\mathbf{x},\mathbf{x}')
        = \sigma_f^2 
        \exp\!\left(-\frac{\lVert \mathbf{x}-\mathbf{x}' \rVert^2}{2\ell^2}\right)
        + \sigma_n^2 \delta_{\mathbf{x}\mathbf{x}'}.
    \end{equation}
\end{itemize}

GPR is a non-parametric method, in which the kernel hyper-parameters are optimized during the regression process. The kernel functions adopted from this work are listed in Table \ref{tab:1}.

\subsection{Neural Networks} 

Multi-layer perceptrons (MLPs) are a class of basic feed-forward neural networks consisting of an input layer, multiple hidden layers, and an output layer, and are widely used for non-linear regression and classification tasks on tabular data \citep{MLP}. Within our multi-model regression framework, MLPs constitute an important family of distinct models.

We construct a suite of independent MLP regression models by systematically varying two core architectural components: the network depth, with models comprising 3, 5, and 7 hidden layers, and the choice of activation function. Specifically, ReLU, Tanh, and Sigmoid activation functions are considered to explore different non-linear response characteristics. All MLP models are implemented using the \texttt{TensorFlow} framework \citep{TensorFlow}. To improve training stability and mitigate over-fitting, batch normalization layers are applied after each dense layer, and dropout regularization with a dropout rate of 0.3 is employed during training. Model optimization is performed using the Adam optimizer. In addition, all models are trained using a robust optimization strategy incorporating early stopping and dynamic learning rate scheduling to promote stable convergence and enhance generalization performance.

\subsection{Additional methods}

This group comprises four deep learning models specifically designed for tabular data: the prior-fitted meta-learning model (TabPFN; \citealt{Tabpfn}), a Transformer-based architecture (FT-Transformer; \citealt{FT-Transformer}), an attention-driven tabular network (TabNet; \citealt{TabNet}), and a differentiable decision tree ensemble (NODE; \citealt{NODE}). These models enhance representation learning and feature interaction beyond conventional multilayer perceptrons while remaining well suited for structured, heterogeneous feature sets. All models in this group are implemented and trained using the \texttt{PyTorch} framework \citep{Pytorch}. Collectively, they provide complementary perspectives on deep learning–based regression for tabular astrophysical data, forming an integral component of our multi-model framework.

\subsection{Model training and testing}

We randomly split the 1,743 blazars with known redshifts into training and test sets with a ratio of 3:1. To assess the robustness of the model comparison with respect to the choice of train/test partition, we repeated this procedure using 100 independent random train/test splits (random seed from 0 to 99), following the validation strategy adopted in recent machine-learning studies of astrophysical redshift estimation (e.g., \citealt{2024ApJ...967L..30D,2024ApJS..271...22D}). For each split, all 35 regression models were retrained from scratch using identical hyperparameter settings and evaluated on the corresponding independent test set.

Model performance is quantified using the mean absolute error (MAE), root mean square error (RMSE), and the Pearson correlation coefficient between the predicted and true redshifts. The performance metrics reported in Table~\ref{tab:1} correspond to the mean values and their associated standard deviations obtained from these 100 independent random train/test splits, providing a statistically more robust comparison of model performance than that based on a single data partition.

\section{Model comparison} \label{sec:comparison}

As shown in Table~\ref{tab:1}, the predictive performances of the 35 regression models are confined within a relatively narrow range. The mean MAE values span only 0.09916--0.1166, while the mean RMSEs range from 0.1291 to 0.1472. Likewise, the Pearson correlation coefficients differ only modestly among the models. TabPFN achieves the best overall performance, whereas the Decision Tree performs worst, mainly because the tree depth is intentionally constrained to mitigate over-fitting, thereby limiting the model capacity. In general, tree-based ensemble methods and modern tabular deep-learning models exhibit slightly better performance than linear and kernel-based approaches, although the differences remain relatively small.

This limited performance spread is also consistent with previous machine-learning studies of AGN and GRB redshift estimation (e.g., \citealt{2021ApJ...920..118D,2022ApJS..259...55N}), where different model architectures and sample-selection strategies likewise converge to comparable prediction accuracies. Despite differences in datasets (e.g., 4LAC-DR2 versus DR3), source-selection criteria, and machine-learning methodologies, the reported prediction errors remain remarkably similar to those obtained in this work.

this suggest that the predictive performance has reached a performance plateau under the current feature. Although increasingly sophisticated models provide modest improvements, the overall gains remain limited, implying that the predictive accuracy is more likely constrained by the information content of the available observational features than by model architecture or capacity. We emphasize that this interpretation is empirical and should be regarded as a plausible explanation rather than a definitive conclusion.

 The repeated random-split analysis demonstrates that the relative performance of the models is largely insensitive to the particular data partition. Therefore, adopting a fixed reference split provides a reproducible and internally consistent training pipeline for all downstream analyses. Unless otherwise stated, all subsequent analyses presented in this paper are based on a single reference train/test split (random seed = 123).The redshift distributions of the training set (upper panel) and test set (bottom panel) for this reference split are shown as solid lines in Figure~\ref{fig:4}. No significant deviation is observed between the two distributions, indicating that the adopted partition preserves the statistical properties of the full sample.

Figure~\ref{fig:2} presents the scatter plots between the predicted and true values of $\rm 1/(1+z)$ for all 35 regression models using the reference train/test split. The black dashed line denotes the ideal one-to-one relation. Most data points cluster around this line, demonstrating that all models successfully recover the global trend of the redshift distribution. The remaining scatter primarily reflects the intrinsic ambiguity in mapping the available observational features to redshift rather than strong systematic biases in any individual model.

\begin{figure}
\centering
  \includegraphics[width=0.48\textwidth]{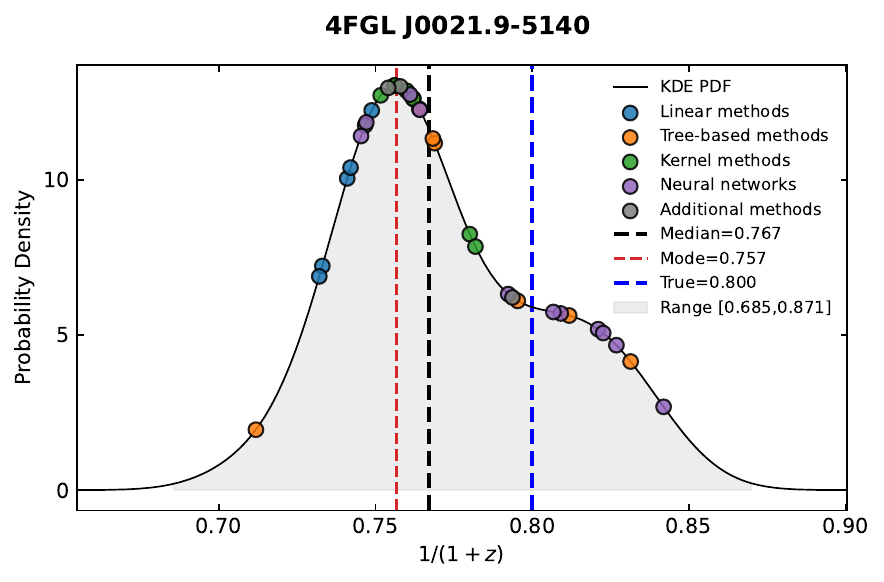}
\caption{
Example of the predictive distribution constructed for a randomly selected test-set source. Independent redshift predictions from different regression models are shown as colored points. The KDE-based PDF is plotted as a black solid curve, with the shaded gray region indicating the $\rm 3\sigma$-equivalent (99.73\%) confidence interval. The median and mode of the predictive distribution are marked by black and red solid curves, respectively, while the true redshift is indicated by a blue dashed line.
}
\label{fig:3}
\end{figure}

Although the overall predictive performances of different models are highly similar, the predictions for an individual source can differ substantially. Figure~\ref{fig:3} illustrates this effect using a representative test-set source. For a source with a true value of $\rm 1/(1+z)=0.80$, the predictions from the 35 independent models span the interval 0.72--0.85. Similar behavior is commonly observed throughout the test sample, indicating that comparable global performance metrics do not necessarily imply consistent predictions at the object level. This dispersion motivates the construction of predictive distributions and uncertainty quantification by combining multiple independent model realizations rather than relying solely on individual point estimates.

\section{Results} \label{sec:results}

\subsection{Redshift distribution of individual blazar} 

For an individual source, the predicted results obtained from the 35 independent regression models are denoted as $\{p_i=1/(1+z_i)\}_{i=1}^{35}$. 
These discrete predictions are treated as samples drawn from an underlying continuous predictive distribution. 
We introduce $p$ as a continuous variable representing the predicted quantity, and interpret the set $\{p_i\}$ within a non-parametric statistical modeling framework. 
Rather than assuming any specific prior distributional form, such as a Gaussian distribution, we interpret these predictions within a non-parametric statistical modeling framework and use kernel density estimation (KDE) to characterize their distribution in range $[0.1, 1.1]$. 
The KDE-based probability density function (PDF) is given by
\begin{equation}
f(p) = \frac{1}{N h} \sum_{i=1}^{N} K\!\left( \frac{p - p_i}{h} \right),
\end{equation}
where $N = 35$ is the number of independent model realizations, $h$ is the bandwidth (automatically selected via Scott’s rule), and $K(\cdot)$ denotes the Gaussian kernel function. 
The resulting PDF is normalized as $\rm \int f(p)\,\mathrm{d}p = 1$. Thus, the probability that the predicted value lies within a finite interval $[p_1, p_2]$ in a continuous predictive distribution is given by
\begin{equation}
\Pr(p \in [p_1, p_2]) = \int_{p_1}^{p_2} f(p)\,\mathrm{d}p .
\end{equation}

From the KDE-based PDF, we further define representative point estimates and uncertainty intervals.
Predictive uncertainty is quantified using probability intervals derived from the KDE-based PDF.
In particular, we adopt the $3\sigma$-equivalent uncertainty interval $[p_{\min},p_{\max}]$, defined as the interval enclosing 99.73\% of the total probability mass,
\begin{equation}
\int_{p_{\rm min}}^{p_{\rm max}} f(p)\,\mathrm{d}p = 0.9973 .
\end{equation}
The mode $p_{\mathrm{mode}}$ is defined as the location of the maximum of the PDF,
\begin{equation}
p_{\mathrm{mode}} = \arg\max_p f(p).
\end{equation}
The median $p_{\mathrm{med}}$ is defined such that
\begin{equation}
\int_{0.1}^{p_{\mathrm{med}}} f(p)\,\mathrm{d}p = 0.5 .
\end{equation}

Figure 3 illustrates the construction of the predictive distribution for a randomly selected source from the test set. Independent redshift predictions from different regression models are shown as colored scatter points. The KDE-based PDF constructed from these predictions is plotted as a black solid curve, with the corresponding $\rm 3\sigma$-equivalent confidence region (i.e., the 99.73\% probability interval) indicated by the gray shaded area. The median and mode of the predictive distribution are marked by black and blue vertical lines, respectively, while the true redshift is shown as a blue dashed line.

By constructing the predictive redshift probability distribution, we effectively integrate the predictions from 35 independent regression models and quantitatively characterize the predictive uncertainty for individual sources. For sources with known redshifts, approximately 65\% of the true values fall within the $\rm 3\sigma$-equivalent  uncertainty interval, exceeding the corresponding fraction obtained from \cite{2024MNRAS.527.6198G}. Meanwhile, the resulting uncertainty intervals are comparable to those of \cite{2024MNRAS.527.6198G}, demonstrating that our approach achieves similar performance while retaining the full sample including faint and low STO sources.

Based on the above procedure, we integrate the predictions from multiple regression models to estimate the redshifts of 1590 blazars with previously missing redshift measurements. For each source, we construct the corresponding predictive distribution and perform uncertainty quantification. The resulting redshift distribution catalog for the full sample is provided in the Appendix.

\subsection{Redshift distributions of blazar subclasses} 

\begin{figure}
\centering
  \includegraphics[width=0.48\textwidth]{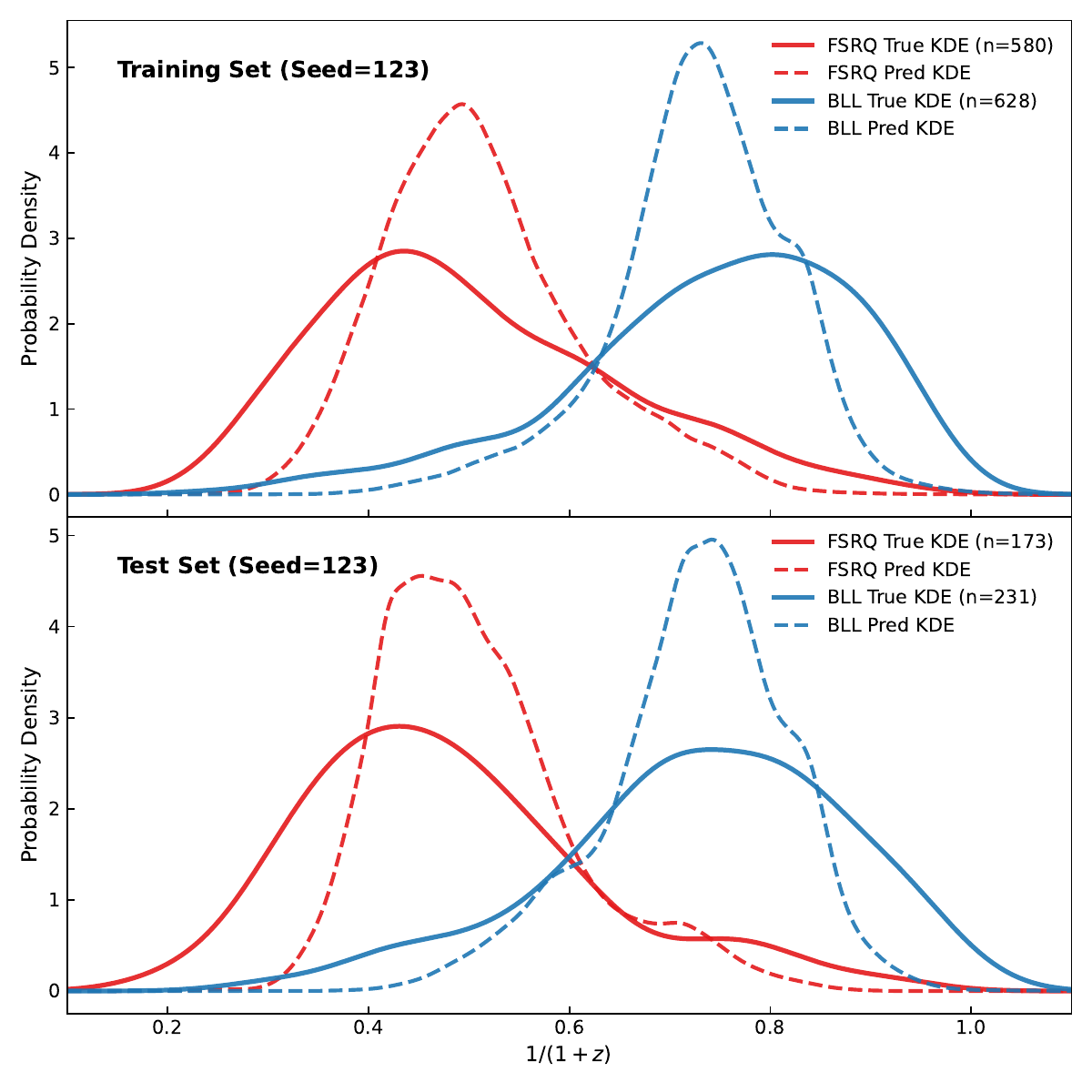}
\caption{
Population redshift distributions of different blazar subclasses in the $\rm 1/(1+z)$ space.
Top panel: distributions for sources in the training set. Bottom panel: distributions for sources in the test set.
In each panel, the solid curves show the distributions of sources with known redshifts for FSRQs (red) and BLLs (blue), while the dashed curves represent the predicted distributions obtained by stacking and averaging the KDE-based PDFs of individual sources.
}
\label{fig:4}
\end{figure}

\begin{figure}
\centering
\includegraphics[width=0.48\textwidth]{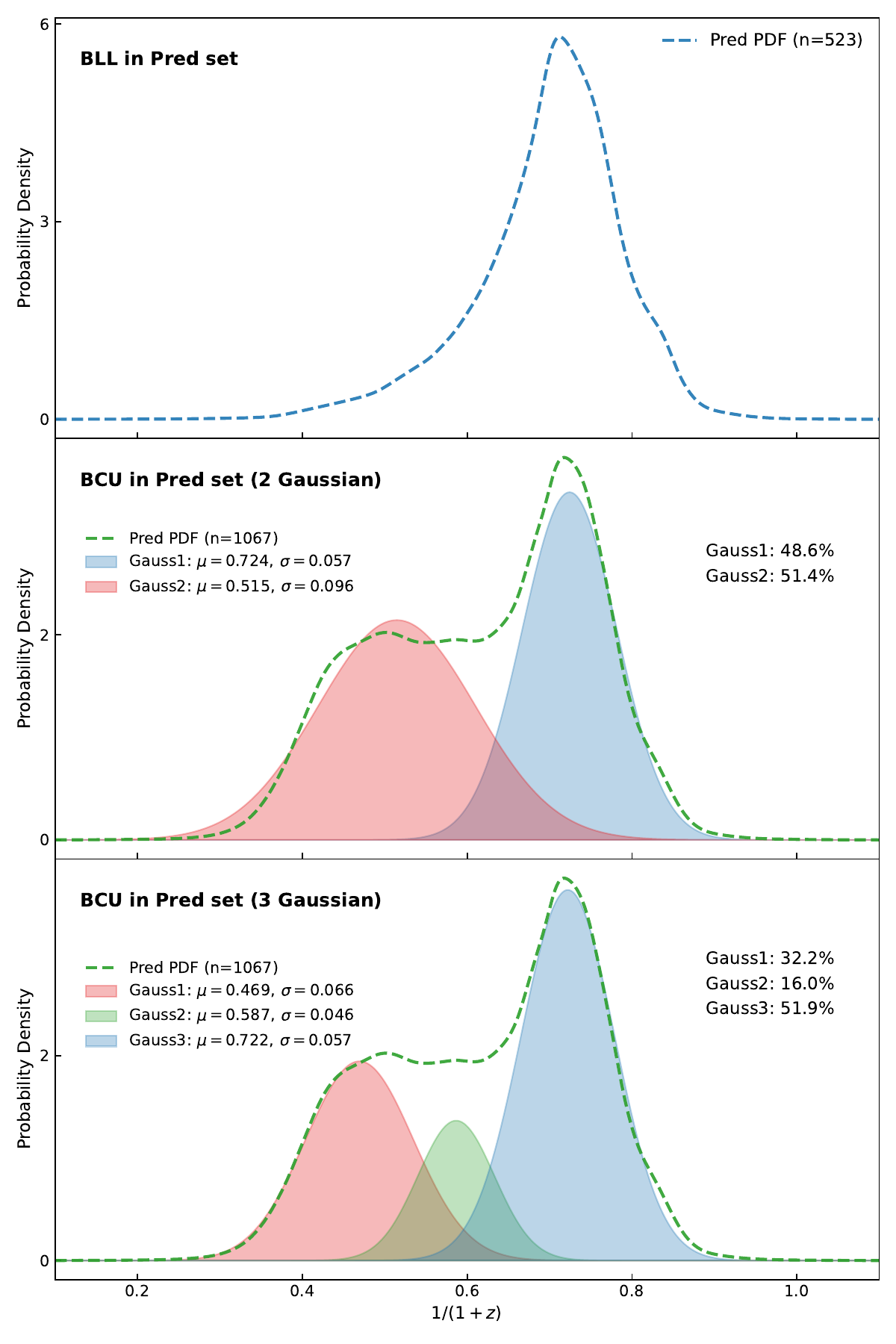}
\caption{
Predicted population redshift distributions for blazars without reported redshifts. The upper panel shows the distribution for BLLs. The middle and lower panels show the BCU distribution fitted with two- and three-component Gaussian mixture models, respectively. 
}
\label{fig:5}
\end{figure}

We next investigate the population redshift distributions of different blazar subclasses. As shown by the solid curves in Figure \ref{fig:4}, the blazar population exhibits a clear double-peaked distribution in the $\rm 1/(1+z)$ space. The peak associated with FSRQs is located at lower $\rm 1/(1+z)$ values, corresponding to higher redshifts, whereas BLLs preferentially populate higher $\rm 1/(1+z)$ values, corresponding to lower redshifts. This likely reflects differences in the blazar evolution sequences of the two subclasses, as well as observational selection effects.

We further construct the predicted population redshift distributions by stacking and averaging the predicted KDE PDFs of individual sources in both the training and test set. The resulting distributions are shown as dashed curves in Figure \ref{fig:4}. The predicted redshift distributions also display a pronounced double-peaked structure, with peak locations closely matching those of the true distributions, albeit with systematically narrower widths. 

It is worth emphasizing that the FSRQ/BLL classification is not included as an input feature in the model training. Nevertheless, the models are able to recover the distinct redshift distribution patterns of the two subclasses based solely on other feature parameters, effectively separating FSRQs and BLLs. This indicates that the adopted feature set implicitly encodes information related to both source class and its characteristic redshift range. On the other hand, the predictive performance degrades at the extremes of the redshift distribution. For sources with $1/(1+z) < 0.3$ (corresponding to $z > 2.33$) and $1/(1+z) > 0.9$ (corresponding to $z < 0.11$), the predicted redshifts tend to regress toward intermediate values and fail to reproduce the extended tails of the true distribution. This behavior primarily reflects the scarcity of training samples at the extreme ends of the redshift distribution, which limits the model's ability to accurately characterize these regions and results in predictions that are drawn toward the data-rich intermediate values.

Figure \ref{fig:5} presents the predicted population redshift distributions for blazars without reported redshifts. The upper panel shows the distribution for BLLs, which exhibits a unimodal structure peaking around $\rm 1/(1+z)\simeq0.7$, broadly consistent with the distribution inferred from BLLs with known redshifts.

The middle and lower panels present the predicted redshift distribution for BCUs together with two- and three-component Gaussian mixture representations, respectively. The inferred BCU distribution exhibits a pronounced multi-peaked structure. The two-component Gaussian mixture provides a satisfactory empirical description of the distribution, yielding two Gaussian components with means and standard deviations of $\rm (\mu_1=0.724,\sigma_1=0.057)$ and $\rm (\mu_2=0.515,\sigma_2=0.096)$, with relative weights of approximately 48.6\% and 51.4\%, respectively. The component centered at $\mu_1=0.724$ corresponds to a mean redshift of $z\simeq0.38$, consistent with the characteristic redshift range of BLLs in the redshift-known sample, whereas the component centered at $\mu_2=0.515$ corresponds to $z\simeq0.94$, close to the typical redshift regime of FSRQs. The two-component representation therefore naturally separates the BCU population into BLL-like and FSRQ-like contributions.

We also fitted the distribution using a three-component Gaussian mixture model, which likewise provides a satisfactory description of the inferred distribution. Compared with the two-component model, the additional Gaussian component mainly represents the intermediate region between the two dominant peaks. Although the physical origin of this component remains uncertain, it may be related to the continuous transition between different blazar subclasses or to the intrinsic diversity of the BCU population. We therefore regard both the two- and three-component Gaussian mixture models as empirical parameterizations of the inferred redshift distribution rather than definitive evidence for distinct physical populations.

Overall, the inferred distribution suggests that FSRQ-like sources may constitute a slightly larger fraction of the BCU population than BLL-like sources. This inference differs from several previous machine-learning classification studies of the 4LAC-DR3 BCU sample, which generally suggested a BLL-dominated population (e.g., \citealt{2023ApJ...946..109A,2023ApJ...956...48X,2023MNRAS.525.1731C,2024MNRAS.528..976B}). However, it is consistent with the analysis of the Fermi-LAT Collaboration, which showed that reproducing the photon-index distribution of BCUs in 4LAC-DR3 requires a larger FSRQ contribution than that inferred from the classified sample \citep{2022ApJS..263...24A}.

\begin{figure*}
\sidecaption
\begin{minipage}{0.70\textwidth}
\centering
\includegraphics[width=0.49\linewidth]{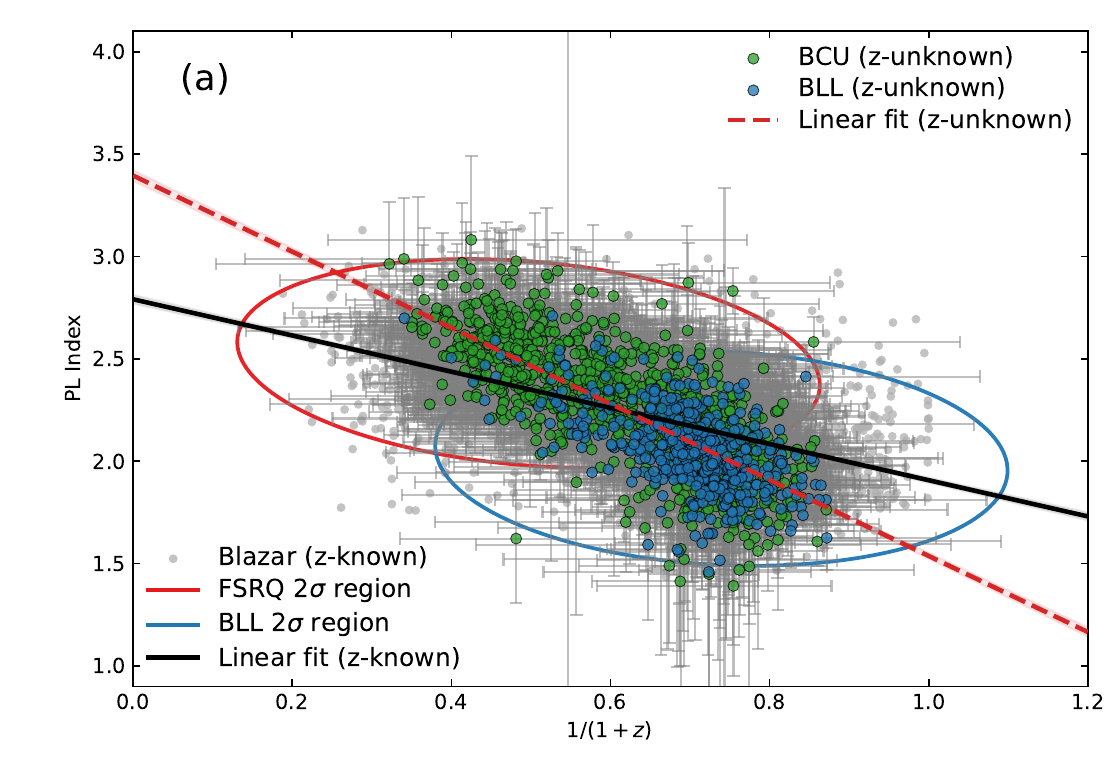}
\hfill
\includegraphics[width=0.49\linewidth]{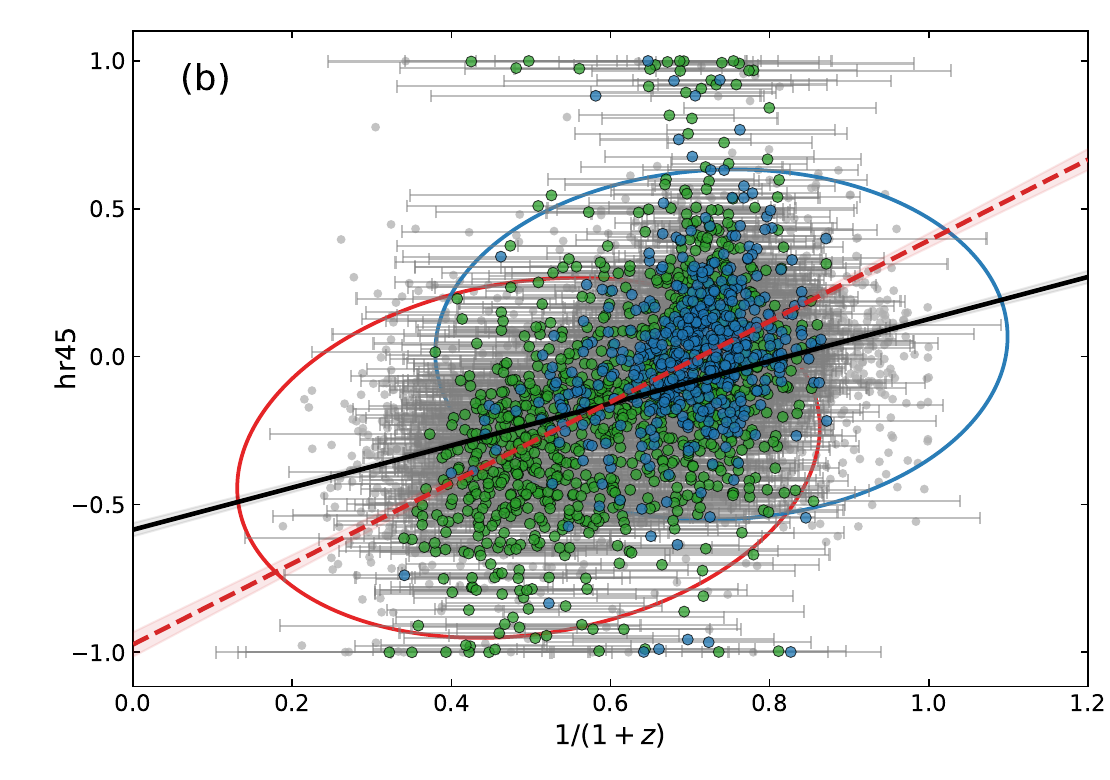}\\[0.5em]
\includegraphics[width=0.49\linewidth]{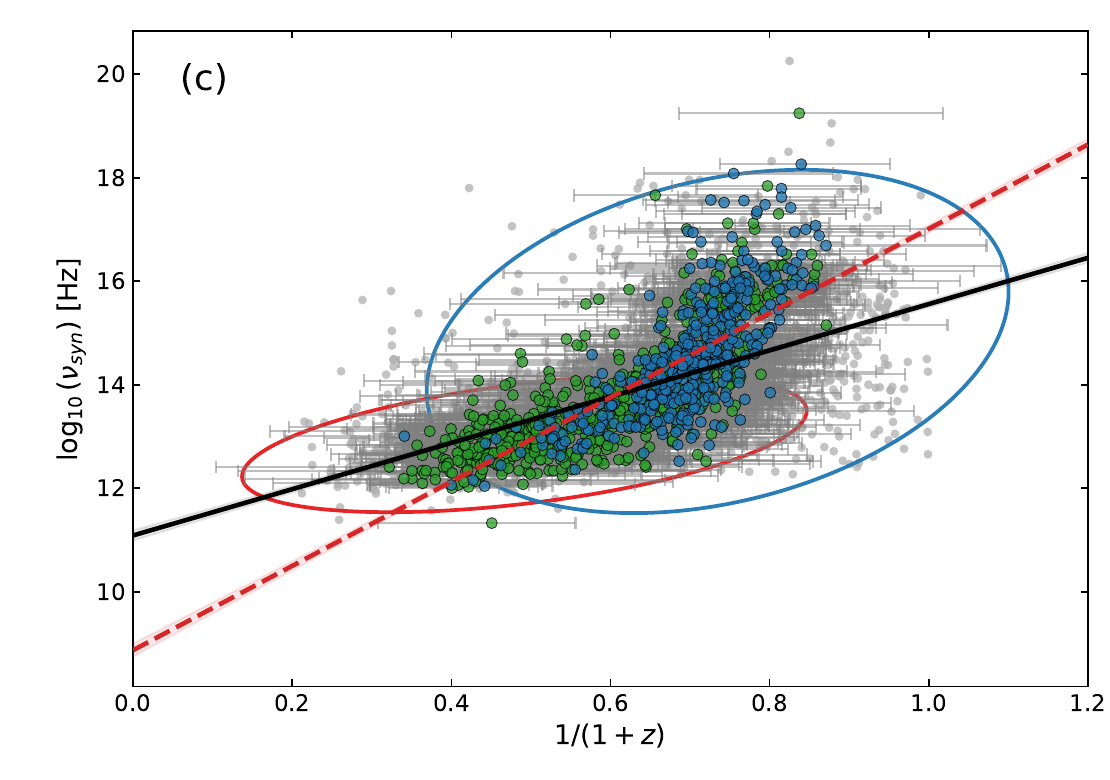}
\hfill
\includegraphics[width=0.49\linewidth]{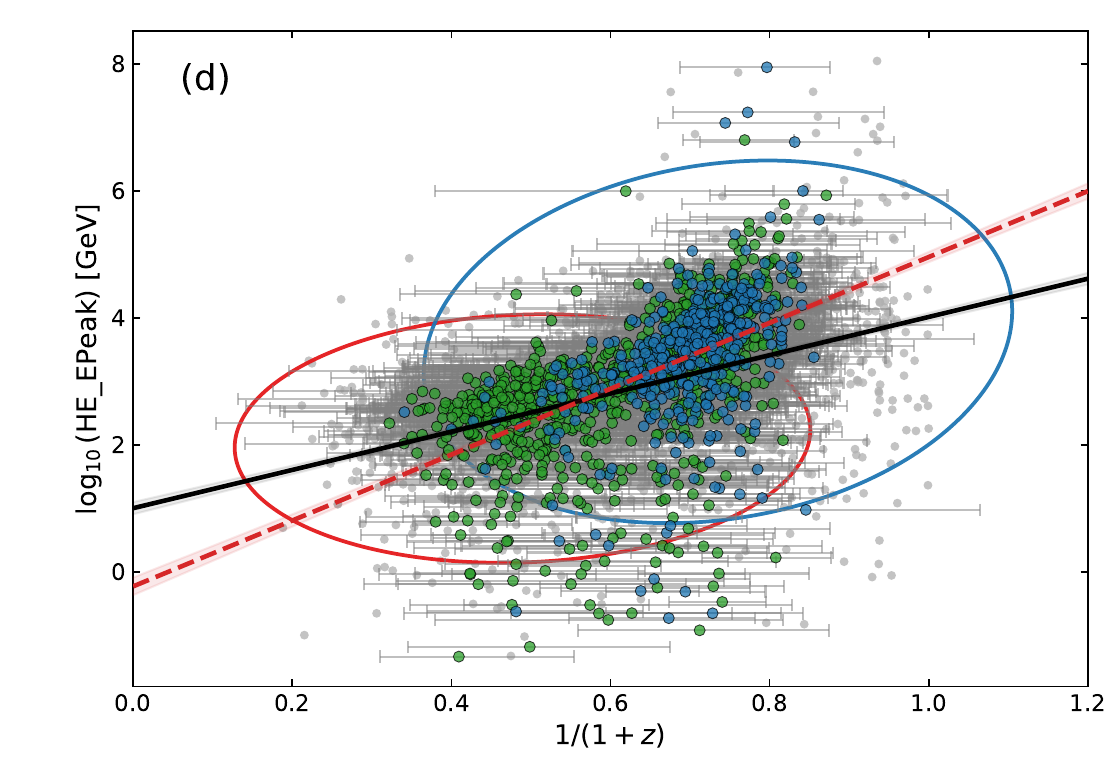}
\end{minipage}
\caption{Parameter--redshift distributions for four representative parameters: (a) PL-Index, (b) $hr_{45}$, (c) $\nu_{syn}$, and (d) HE-EPeak. Gray points denote blazars with measured redshifts, with black solid lines showing the corresponding linear fits. Red and blue ellipses indicate the $2\sigma$ confidence regions of FSRQs and BLLs inferred from Gaussian mixture models. Blue and green points represent the predicted results for redshift-unknown BLLs and BCUs, respectively, while red dashed lines show the linear fits to the redshift-unknown samples.}
\label{fig:6}
\end{figure*}

To assess the reasonableness of the predicted results, we examine the distribution of the predicted redshifts in the parameter space. As shown in Figure \ref{fig:6}, we select four parameters, (a) PL-Index, (b) $hr_{45}$, (c) $\nu_{syn}$, and (d) HE-EPeak, which exhibit the strongest correlations with redshift, and construct their distributions as a function of redshift.
All blazars with reported redshifts are shown as background samples and are represented by gray points. The linear relations between the true redshifts and the corresponding parameters are indicated by black solid lines. Using a Gaussian mixture model (GMM), we further characterize the two-dimensional distributions of FSRQs and BLLs in the parameter space, which are illustrated by red and blue $2\sigma$ confidence ellipses, respectively.
For sources without measured redshifts, the predicted results for BLLs are shown as blue points, while those for BCUs are shown as green points. The linear relations between the predicted redshifts and the corresponding parameters are indicated by red dashed lines.

As can be seen from Figure \ref{fig:6}, the predicted redshifts of BLLs predominantly fall within the $2\sigma$ confidence region of the BLL population. In contrast, the predicted redshifts of BCUs are distributed across both the BLL and FSRQ $2\sigma$ regions, with approximately comparable fractions in the two regions (roughly 1:1). No significant anomalies or inconsistencies relative to the distributions of redshift-known sources are observed.
We further note that the distributions of the redshift-unknown sources appear more concentrated than those of the redshift-known samples. This behavior is likely related to the limited predictive power of the regression models for extreme redshift values. As a consequence, the regression slopes for the redshift-unknown samples are steeper than those derived from the redshift-known sources (red dashed lines versus black solid lines), indicating the potential need for bias correction based on linear reconstruction (see discussion).

\section{Conclusion and discussion} \label{sec:discussion}

We present an integrated framework for redshift distribution estimation using multiple machine-learning models.
Using KNN interpolation to handle missing values, we retain the full blazar sample at high Galactic latitudes in the 4LAC-DR3 catalog and construct a feature set consisting of 40 parameters. A total of 35 independent regression models are trained and tested within a unified process. Predictions from different models are treated as independent redshift estimates, from which source-level redshift distribution are constructed to characterize both its expected value and associated uncertainty. We compile a redshift distribution catalog for 1,590 blazars without measured redshifts. By stacking the source-level PDFs, we further reconstruct population redshift distributions.
The main results are summarized as follows:
\begin{itemize}
\item[1.]
Across a wide range of regression algorithms, including linear models, tree-based methods, kernel methods, neural networks, and deep tabular models, the predictive performance reaches a comparable level, with a $\mathrm{RMSE}\sim 0.1373$, and Pearson correlation coefficients around $\sim0.7$. This indicates the presence of a performance plateau driven primarily by the limited redshift-related information available in the current feature set, rather than by the specific choice of regression model.
\item[2.]
By treating predictions from multiple independent models as empirical samples, we construct non-parametric redshift probability density functions (PDFs) for individual sources, which allow us to characterize predictive uncertainty.
We further provide a catalog of redshift distributions for 1,590 blazars without measured redshifts, including the representative point estimates and uncertainty intervals.
\item[3.]
Stacking the source-level PDFs successfully reconstructs the population redshift distributions for different blazar subclasses. Without using class labels as input features, the predicted distributions not only recover the characteristic double-peaked structure observed in the known-redshift sample but also effectively separate the FSRQ-like and BLL-like populations. For BCUs, the double-peaked structure of the predicted redshift distribution indicates a slightly greater contribution from FSRQ-like sources than from BLL-like ones, a finding that aligns with the indication from the photon-index analysis in 4LAC-DR3.
\end{itemize}

Ensembling the results of multiple machine-learning models has long been an important research topic. Widely adopted ensemble strategies include union-based combinations (e.g., \citealt{2016ApJ...820....8S,2021RAA....21...15Z,2022A&A...660A..87B,2023ApJ...946..109A}), voting schemes (e.g., \citealt{2024MNRAS.527.1794Z}), and more sophisticated meta-algorithms such as SuperLearner (e.g., \citealt{2019arXiv190705074D,2021ApJ...920..118D,2022ApJS..259...55N,2025A&A...698A..92N}), which use cross-validation to optimally combine predictions from a diverse model library.

Benefiting from rapid methodological advances, a broad spectrum of algorithms has become increasingly accessible, enabling the practical deployment of heterogeneous ensemble frameworks. Because different algorithms possess distinct inductive biases and learning strategies, they may provide different yet plausible mappings between observational features and redshift. Unlike methods such as SuperLearner, whose objective is to optimize a single point prediction through weighted model aggregation, our framework deliberately preserves the diversity of predictions from heterogeneous models. This diversity contains useful information about the variability associated with model choice, which reflects the degree of constraint imposed by the training data, with broader scatter empirically flagging regions of weaker information and provides a data-driven basis for assessing the reliability of the inferred redshift estimates.

In this work, we construct redshift PDFs from the ensemble predictions using KDE. The resulting PDFs should be interpreted as empirical ensemble-based predictive distributions, which summarize the variability among different machine-learning models under the adopted modeling framework. Rather than representing a single optimal estimate, they provide a data-driven characterization of the plausible range of redshift estimates supported by the ensemble. They are not intended to represent formally calibrated Bayesian posterior distributions or a rigorous decomposition of epistemic and aleatoric uncertainties.

In astronomical data sets, samples with known labels are generally associated with brighter sources with higher signal-to-noise ratios, for which physical parameters and multi-wavelength observations are relatively complete. 
In contrast, samples requiring machine-learning–based prediction are typically fainter, more weakly constrained observationally, and often suffer from missing or incomplete features. 
This systematic discrepancy, driven by observational selection effects, poses a significant challenge to the representativeness of training samples in supervised learning. 
Although aggressive sample cleaning may improve the overall signal-to-noise ratio of the training data and lead to enhanced predictive performance, it can also reduce the representativeness of the sample and compromise the model’s ability to generalize to the full population of faint sources.
Consequently, how to properly treat the limited number of faint sources in the training set, whose physical properties more closely resemble those of the prediction targets, and how to incorporate physically meaningful constraints into the models are critical for improving generalization to the prediction sample.
In practice, maintaining low STO sources within the modeling framework is necessary to ensure representativeness and applicability to the full survey population.

Previous studies have systematically evaluated the importance of individual features in redshift prediction models and found that source class is among the most influential parameters (e.g., \citealt{2023MNRAS.521.4156C}). 
We have verified that including source class as an input feature can indeed improve test-set performance, improving both the RMSE and the correlation coefficient by approximately 7.5\%. 
However, in the data set considered here, the redshift-known sample is dominated by FSRQs and BLLs, with BCUs accounting for only a small fraction ($\sim$7.5\%), whereas the redshift-unknown sample consists primarily of BCUs, with a minor contribution from BLLs and no FSRQs. 
Given this mismatch in class composition, incorporating source class as an input feature is not statistically appropriate for the target population and may introduce systematic biases, while beneficial for test-set performance. 
Consequently, the source class is not included as an input feature in the present work.

A bias of this study lies in its reduced sensitivity to extreme redshift values. 
Although modeling in the $\rm 1/(1+z)$ space alleviates the strong skewness of the redshift distribution, this transformation also compresses the separation between very-low and very-high redshift regimes. 
Due to the lack of extreme-redshift sources, the models are weakly constrained in these regions, causing predictions to 
drift toward intermediate values and leading to the formation of relatively sharp peaks in the inferred distributions, consistent with the tendency of regression models to revert toward the mean in poorly constrained regions. 
As shown in Figure~\ref{fig:6}, the limited predictive capability at extreme redshifts leads to a systematic bias in which the predicted redshift values are shifted toward intermediate values, causing the predictions to cluster around the mean and distorting the observed photon-index--redshift relation.

To mitigate this effect, introducing additional constraints for distribution reconstruction represents a promising approach. Two main strategies can be considered. 
(1) Empirical calibration based on predicted values. \cite{2021ApJ...920..118D} proposed a class-based linear reconstruction method that exploits the linear relationship between true redshifts and model predictions in the training sample to systematically correct biases for specific source classes. 
(2) The imposition of physically motivated constraints based on feature--redshift relations. In particular, observable quantities such as the photon index and peaked frequency exhibit approximately linear correlations with redshift (as shown in Figure \ref{fig:3}), which can be leveraged to perform linear reconstruction of source redshift distributions. This approach does not rely on source class labels, but instead directly utilizes physically meaningful observables. 
The effective integration of prior physical knowledge into the prediction framework, with the aim of improving inference for extreme-redshift and observationally incomplete sources, will be explored in future work.

\section*{Data availability}\label{sec:dataavai}

The machine-readable catalog described in Appendix~A is available only in electronic form at the CDS via anonymous FTP to \texttt{cdsarc.u-strasbg.fr} (130.79.128.5) or via \url{http://cdsweb.u-strasbg.fr/cgi-bin/qcat?J/A+A/}. The catalog contains the predictions from the 35 individual regression models, the ensemble median and mode estimates, the corresponding uncertainty intervals, and the source-level probability density functions for the 1\,590 blazars without spectroscopic redshift measurements. The complete set of source-level redshift probability distribution plots is publicly available on Zenodo at \url{https://doi.org/10.5281/zenodo.22875440}.

\section*{Acknowledgements}
We thank the anonymous referee for very constructive and helpful comments and suggestions, which greatly helped us to improve our paper. 
This work is supported by the National Natural Science Foundation of China (NSFC; Grant Nos.~12603160 and 12233006). 
Dr.~Shan Chang is supported by the Xingdian Talent Support Plan -- Youth Project. 

This work made use of publicly available data and catalog products from the Fermi Large Area Telescope (Fermi-LAT), including the 4LAC-DR3 and 4FGL-DR4 catalogs, accessed through the Fermi Science Support Center (FSSC). We acknowledge the Fermi-LAT Collaboration for making these resources available to the scientific community.

{\it Software:} Scikit-learn \citep{Scikit-learn}, SciPy \citep{SciPy}, Astropy \citep{Astropy}, Matplotlib \citep{Matplotlib}, XGBoost \citep{XgBoost}, LightGBM \citep{LightGBM}, CatBoost \citep{CatBoost}, NumPy \citep{NumPy}, TensorFlow \citep{TensorFlow}, PyTorch \citep{Pytorch}, TabPFN (version~2.0; \citealt{Tabpfn}).

\bibliographystyle{aa}
\bibliography{ref}

\clearpage

\begin{appendix} 
\section{Blazar redshift distribution catalog}

Following the analysis described above, we compile predicted redshift results for 1,590 blazars without spectroscopic measurements. The released data consist of two components: (i) source-level redshift probability distribution plots and (ii) tabulated data containing the predictions from individual models as well as the ensemble-based results, including the corresponding uncertainty estimates.

\begin{figure*}
\centering
\includegraphics[width=\textwidth]{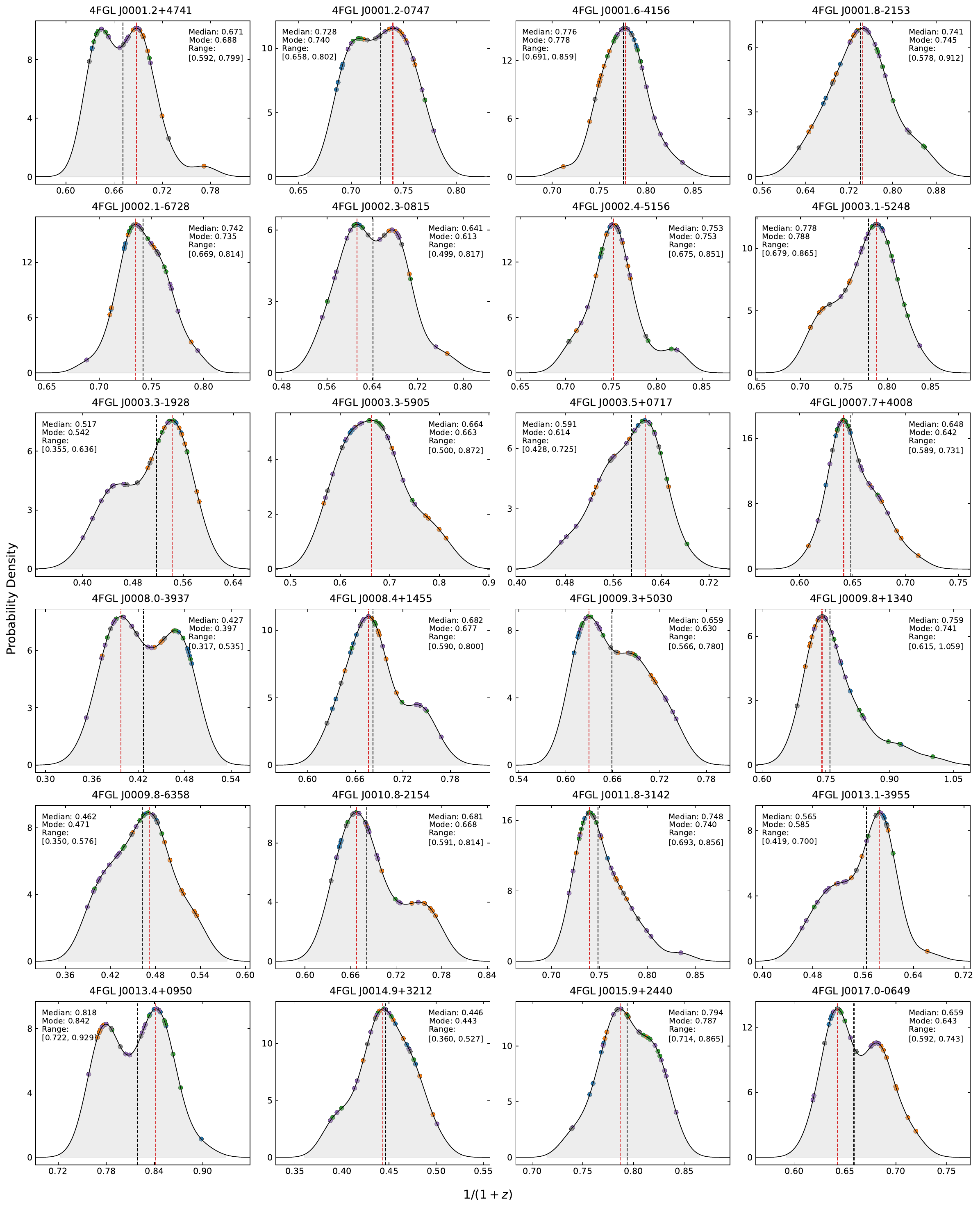}
\caption{Predicted redshift probability distributions for the first 20 blazars without measured redshifts. Colored points represent the independent predictions from the 35 regression models. The black solid curve denotes the KDE-based probability density function. The black and blue dashed vertical lines indicate the predicted median and mode values, respectively. The gray shaded region marks the $3\sigma$-equivalent confidence interval. The legend follows the same convention as in Fig. 3, except that no true redshift values are shown.}
 \label{figA}
\end{figure*}

Figure \ref{figA} shows the predicted redshift probability distributions for the first 20 sources without measured redshifts. The legend is identical to that in Fig. \ref{fig:3}, except that true redshift values are not included. Colored points represent the independent predictions from different regression models. The black solid curve denotes the constructed KDE-based PDF. The black dashed and blue dashed vertical lines indicate the predicted median and mode values, respectively, while the gray shaded region marks the $3\sigma$-equivalent confidence interval. The text annotations in the figure summarize the median, mode, and corresponding confidence interval of the predicted redshift distribution.

For clarity, only the first 20 examples are displayed. The full set of redshift distribution plots for all sources, along with a comprehensive catalog including individual model predictions, ensemble estimates, and associated uncertainty measures, is made available online in FITS format (see the Data availability section).

\end{appendix}

\nolinenumbers

\end{document}